\documentclass[aip,reprint]{revtex4-1}
\usepackage[utf8]{inputenc}
\usepackage{amsmath}
\usepackage{xfrac}
\usepackage{graphicx}
\usepackage{ulem}
\usepackage{xcolor}

\usepackage{siunitx}

\newcommand{\He}[1]{{\(^{#1}\)He}}

\DeclareSIUnit{\sqrthz}{\ensuremath{\sqrt{\text{\hertz}}}}
\DeclareSIUnit{\FluxQuantum}{\ensuremath{\Phi_0}}
\DeclareSIUnit{\nmnoise}{\nano\metre\per\sqrthz}

\begin{document}
\title{Scanning SQUID microscopy in a cryogen-free dilution refrigerator}
\author{D. Low}
\author{G. M. Ferguson}
\author{Alexander Jarjour}
\author{Brian T. Schaefer}
\affiliation{Laboratory of Atomic and Solid-State Physics, Cornell University, Ithaca, NY 14853, USA}
\author{Maja D. Bachmann}
\affiliation{Max Planck Institute for Chemical Physics of Solids, D-01187 Dresden, Germany}
\affiliation{School of Physics and Astronomy, University of St. Andrews, St. Andrews KY16 9SS, UK}
\author{Philip J. W. Moll}
\affiliation{Laboratory of Quantum Materials (QMAT), Institute of Materials, \'{E}cole Polytechnique F\'{e}d\'{e}ral de Lausanne (EPFL), 1015 Lausanne, Switzerland}
\author{Katja C. Nowack}
\email{kcn34@cornell.edu}
\affiliation{Laboratory of Atomic and Solid-State Physics, Cornell University, Ithaca, NY 14853, USA}
\affiliation{Kavli Institute at Cornell for Nanoscale Science, Cornell University, Ithaca, NY 14853, USA}

\date{\today}

\begin{abstract}
    We report a scanning superconducting quantum interference device (SQUID) microscope in a cryogen-free dilution refrigerator with a base temperature at the sample stage of at least \SI{30}{\milli\kelvin}. The microscope is rigidly mounted to the mixing chamber plate to optimize thermal anchoring of the sample. The microscope housing fits into the bore of a superconducting vector magnet, and our design accommodates a large number of wires connecting the sample and sensor. Through a combination of vibration isolation in the cryostat and a rigid microscope housing, we achieve relative vibrations between the SQUID and sample that allow us to image with micrometer resolution over a \SI{150}{\micro\metre} range while the sample stage temperature remains at base temperature. To demonstrate the capabilities of our system, we show images acquired simultaneously of the static magnetic field, magnetic susceptibility, and magnetic fields produced by a current above a superconducting micrometer-scale device.
\end{abstract}

\maketitle

\section{Introduction}
Superconducting Quantum Interference devices (SQUIDs) are among the most sensitive magnetic sensors available and have been widely used for magnetic imaging at cryogenic temperatures. Scanning SQUID microscopy can be used to image static stray magnetic fields above a sample, to measure the local magnetic response and to image the magnetic field produced by applied and spontaneous currents in a device. This technique has been applied to a wide variety of quantum materials and mesoscopic devices including unconventional superconductors \cite{KirtleyReview2010}, topological insulators \cite{NowackNatMat2013,SpantonPRL2014}, complex oxides \cite{BertNatPhys2011, KaliskyNatMat2013, Wang2015ImagingHeterostructures}, superconducting and normal metal rings \cite{KirtleyNatPhys2006, KoshnickScience2007,Frolov2008Imaging-junctions, BluhmPRL2009}, and unconventional Josephson junctions \cite{SochnikovNanoLett2013,SpantonNatPhys2017}. Some physical phenomena require cooling to sub-Kelvin temperatures, which can be achieved in a dilution refrigerator (DR).  In a traditional DR, an inner vacuum chamber is immersed in a liquid \He{4} bath. Within the vacuum chamber, a smaller bath of liquid \He{4} is pumped to lower its temperature to \(\sim\)\SI{1.5}{K}. This bath precools a circulating \He{4}/\He{3}  mixture that ultimately provides continuous cooling to a few millikelvin. In a cryogen-free DR, all liquid \He{4} baths are  replaced by a cryocooler, often a pulse tube, which provides cooling to approximately \SI{3}{K}. Closed-cycle circulation of a \He{4}/\He{3} mixture is still used in these systems to achieve temperatures of a few millikelvin at the mixing chamber.

A cryogen-free DR is attractive due to rising prices and uncertain supply of liquid helium. In addition, no \He{4} transfers are necessary, which can interrupt measurements and prevent running experiments remotely over extended periods of time. Cryogen-free DRs are often built with a large sample volume as they do not require a \He{4} bath to envelop the inner vacuum chamber. However, vibrations from the cryocooler put cryogen-free DRs at a major disadvantage. During the pulse tube cooling cycle, high-pressure helium gas is pushed in and out of the pulse tube at a frequency of approximately 1 Hz, rendering scanning probe microscopy in a cryogen-free DR challenging. To date only a few scanning probe microscopes have been reported that operate in cryogen-free DRs, including a scanning gate microscope  \cite{PelliccioneRSI2013} and a scanning tunneling microscope \cite{denHaanRSI2014}. In both cases, custom spring stages were used to mechanically isolate the microscope from the cryostat.

Scanning SQUID microscopy has been implemented in two different types of cryogen-free cryostats with  base temperatures of \(\sim\)3--4 K, but not in a cryostat reaching lower temperatures. In Ref. \cite{ShperberRSI2019}, the authors used a cryostat with built-in mechanical isolation between the cryocooler and the cold plate on which the microscope is mounted. In Ref. \cite{BishopVanRSI2019}, the microscope was placed in a pulse tube based cryostat similar to our own but without the dilution unit and associated cold plates. A custom spring stage was used to significantly reduce vibrations in the microscope. However, such mechanical isolation often comes at the cost of reduced thermal anchoring: a less rigid mechanical connection implies less thermal conductivity between the microscope and the cold plate. A weak thermal connection can prevent rapid measurements over a large scan window due to the heat piezoelectric elements generate while moving. This is particularly challenging when operating near the base temperature in a DR due to the limited cooling power. In addition, a spring stage is difficult to combine with mechanically rigid low-loss coaxial connections to either the SQUID or the sample. Such connections are necessary to implement a dispersive SQUID readout \cite{HatridgePRB2011, ForoughiAPL2018, LevensonFalkSST2016} or to deliver gigahertz and fast rise time excitations to a sample. Combining a spring stage with a superconducting magnet is challenging as well, since a strong magnetic field may exert forces on the microscope and causes undesirable motion\cite{Pelliccione2013DesignRefrigerator}. In principle, scanning probe microscopes are only affected by relative motion of the probe and the sample. This suggests that a rigid microscope in which the probe and sample move together in response to the pulse-tube-induced vibrations offers an alternative to using spring stages. 

Here, we report the implementation of a scanning SQUID microscope in a cryogen-free DR with a base temperature of \(\sim\)\SI{10}{\milli\kelvin}. The microscope is operated in the bore of a superconducting vector magnet. We avoid the use of a spring stage in order to optimize for low sample temperatures. To achieve an acceptable level of vibrations, we designed the microscope prioritizing rigidity while still maintaining a scan range of \SI[product-units=power]{150x150x100}{\um} and \SI[product-units=power]{6x6x6}{\mm} coarse positioning range.

\section{Description of the scanning SQUID microscope}

\begin{figure}[btp]
    \centering
    \includegraphics{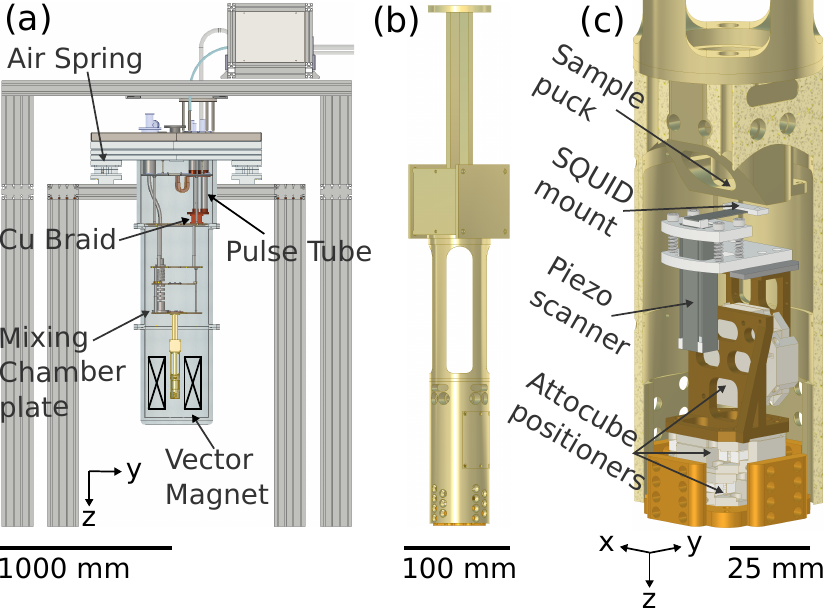}
    \caption{ Drawings of experimental setup and scanning probe microscope. (a) Cross-sectional view of dilution refrigerator with scanning probe microscope.  The room temperature vacuum can is shown. Light tight cans at \SI{60}{K}, \SI{4}{K}, and \SI{1}{K} and various cross beams on both frames are omitted for clarity. (b) Cold finger and microscope housing that attach to the mixing chamber plate. (c) Microscope housing cutaway showing coarse positioners, the piezoelectric scanner, and sample puck. The piezoelectric benders are shown in dark grey. The sample is mounted face-down on the sample puck. The SQUID is mounted on a printed circuit board (not shown) inserted on the end of the horizontal piezoelectric bender providing vertical motion.}
    \label{fig:cad}
\end{figure}

Fig.~\ref{fig:cad} shows drawings of the scanning SQUID microscope mounted in the DR. Computer-aided design drawings of the microscope as shown in (b,c) are available at \cite{CADDrawing}. We use a Bluefors BF-LD 400 DR with a base temperature of \SI{10}{\milli\kelvin} and \SI{400}{\micro W} cooling power at \SI{100}{\milli\kelvin}. Cooling to \(\sim\)\SI{3}{K} is provided by a Cryomech PT-415 pulse tube cooler implemented with a remote motor option. A superconducting vector magnet from American Magnetics Inc.~can apply \SI{6}{T} perpendicular to and \SI{1}{T} in any in-plane direction of the sample.  A few modifications designed and implemented by Bluefors reduce the vibrations of the mixing chamber plate. The DR rests on two nested aluminum frames to reduce the mechanical coupling between the pulse tube cooler and the interior of the cryostat. All cold plates, vacuum cans, and the superconducting magnet are supported by the inner frame while the pulse tube and related components are mounted onto the outer frame. The two nested frames are only connected via an edge-welded stainless steel bellows to maintain vacuum and via copper braids inside the vacuum chamber to thermally link the cold plates and the pulse tube. The top plate of the cryostat is bolted to heavy metal plates (approx. 400 kg). These metal plates rest on air springs mounted to the inner frame. The remote motor head of the pulse tube, which is a source of vibrations, is housed in a sound isolation box that is mounted on the outer frame and connected to the pulse tube by high- and low-pressure flexible lines. We implemented a linear motor driver for the cold head motor as suggested in Ref.\cite{PelliccioneRSI2013}. The high-pressure lines connecting the remote motor head and the compressor are run through thick fiberglass sleeving to reduce the acoustic noise in the lab. The still pump line is passed through a concrete block between the cryostat and the gas handling cabinet. The gas handling cabinet and the compressor are in a utility space that is well separated from the lab space.

The microscope consists of a cold finger and the microscope housing. Each is machined from a single block of copper and bolted together on machined flat, keyed surfaces using four brass screws. The cold finger is mounted to the mixing chamber plate using six brass screws. The outer diameter of the housing is set by the 68 mm diameter bore of the superconducting magnet. The length of the cold finger is chosen such that the sample is positioned at the center of the magnet. Windows in the microscope housing (see Fig.~1b) can be opened by removing copper plates to get visual access when aligning the SQUID to the sample at room temperature.

All wiring runs along the cold finger and enters the microscope housing through a box at the bottom of the cold finger. This box has two removable plates, which allows us to include and change desired wiring feedthroughs. A light tight gold plated copper can (not shown) is bolted below the wiring feedthrough box on the cold finger. This design allows to in principle route fully filtered and shielded wiring to the microscope. All machined parts are gold-plated.

The SQUID is mounted on a home-built piezoelectric scanner similar to the ones described in Refs.\cite{ShperberRSI2019, BishopVanRSI2019}, which in turn is mounted on a three-axis stack of attocube coarse positioners (see Fig.~1c). We chose bearing-based attocube ANPx311 positioners with \SI{6}{mm} coarse range. Two machined brackets adapt a high-load ANPx311 positioner to act as a $z$ positioner. The attocube positioners are made of grade 2 (99\% pure) titanium. The scanner is assembled from piezoelectric bimorphs and machined Macor parts \cite{SiegelRSI1995,BjornssonRSI2001}.  Two pairs of ``s-benders'' move the SQUID in a horizontal plane, while an additional cantilever piezo allows for vertical motion. Imaging range and stiffness compete in the scanner design, since both depend on the dimensions of the benders. Here we chose \(x,y\) s-benders with lateral dimensions \(38.1 \times 6.4\) mm and \SI{0.13}{\milli\metre} thickness. The \(z\) bender has \(25.4 \times 6.35\) mm lateral dimensions and is  \SI{0.19}{\milli\metre} thick. The benders are joined by Macor blocks using EPO-TEK H70e epoxy. The scanner is  attached to a top Macor plate visible in Fig.~\ref{fig:cad}. This plate is mounted with spring-loaded screws to a bottom Macor plate, which in turn is rigidly mounted on the coarse positioners. This arrangement allows us to adjust the alignment angles between the SQUID and the sample through adjusting the spring-loaded screws. Our scan range at cryogenic temperatures is approximately \SI{150}{\micro\metre} in the \(x,y\)-directions and \SI{110}{\micro\metre} in the \(z\)-direction. The scanner and attocube stack are mounted on a copper puck that slides out at the bottom of the microscope housing allowing for easy SQUID replacement. The copper puck is threaded and bolted on two adjacent sides to the microscope housing with four brass screws on each side. To detect when the SQUID touches down on the sample, the SQUID is mounted on a flexible brass cantilever. We monitor the capacitance between this cantilever and a ground plane on the printed circuit board. At touchdown, a sharp increase in the capacitance is detected. These cantilevers are typically \SI{5}{\milli\metre} long, \SI{2}{\milli\metre} wide and \SI{50}{\micro\metre} thick.

The sample is mounted upside-down on a puck that slides into the top of the microscope housing to allow easy sample replacement and is fastened using four brass screws. A ruthenium oxide thermometer is mounted on the sample puck. The base temperature on the mixing chamber plate is \(\sim\)\SI{10}{\milli\kelvin}. The sample mount reaches \SI{50}{\milli\kelvin} at the same time as the mixing chamber plate and continues to cool as the mixing chamber plate cools further. The sample mount thermometer is calibrated to \SI{50}{\milli\kelvin}, but based on the continued change in its resistance, we estimate that the sample mount cools to at least \SI{30}{\milli\kelvin} and likely lower. Importantly, no appreciable heating is observed during scanning at a rate of \(\sim\)\SI[per-mode=symbol]{20}{\micro\metre\per\second} even at base temperature. 

Three wire bundles with 12 twisted pairs are connected to the microscope (two bundles with 36 AWG phosphor bronze, one bundle with 36 AWG Cu and NbTi/CuN wires for low resistance connections). Two of these are used to operate the piezo scanner, the attocube coarse positioners and the SQUID. One bundle is available for sample connections. In addition, the cryostat includes rigid and semi-rigid coaxial wiring to the mixing chamber plate. The combination of 24 wires available to connect to the sample and large coarse positioning range allows us to image devices with several contacts and electrostatic gates, perform full transport characterization of these devices, and image multiple devices in a single cooldown. 
\begin{figure}[btp]
    \centering
    \includegraphics[scale=0.75]{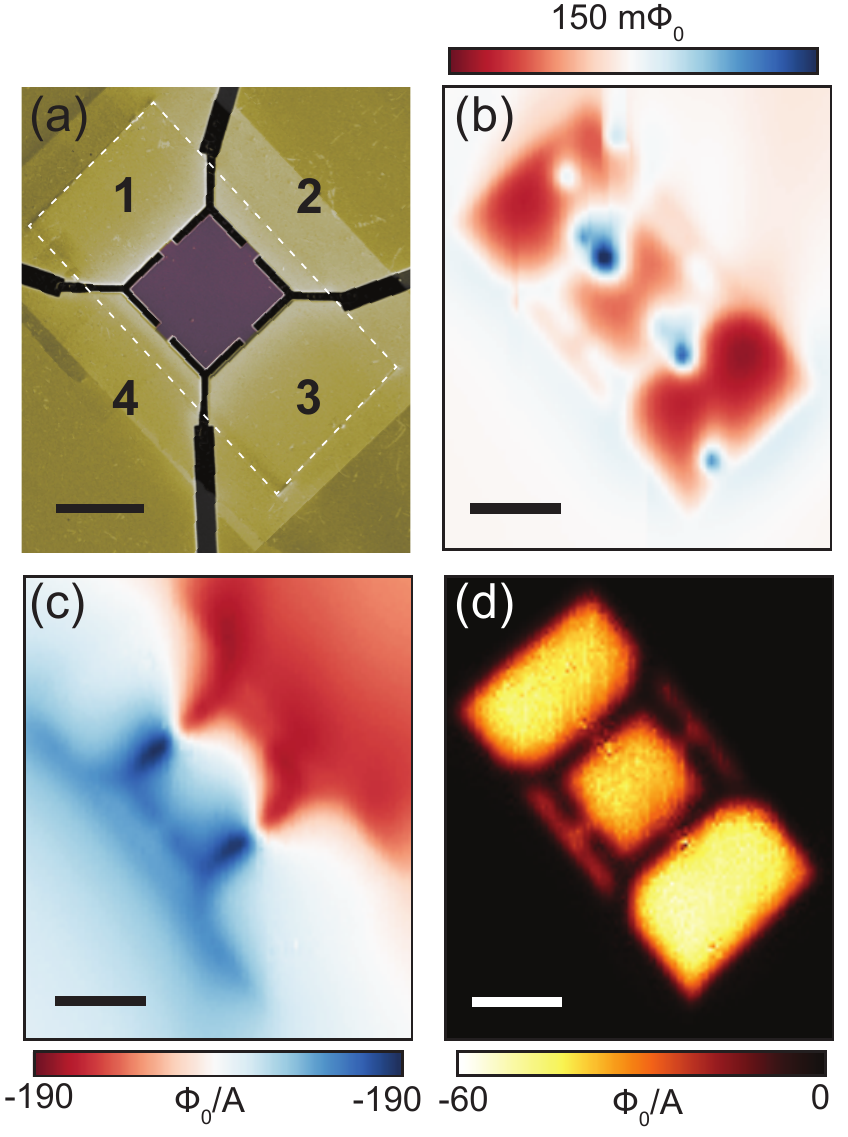}
    \caption{ Scanning SQUID images of a superconducting CeIrIn\(_5\) microstructure fabricated using a focused ion beam. 
    (a) Scanning electron microscope (SEM) image of the device with contacts 1-4.  
    (b) Image of the stray magnetic field above the device. The image shows a combination of Meissner screening (red) of a small background field and and vortices (blue) in the device. 
    (c) Image of the magnetic field produced by a total current of $\sim$ \SI{56}{uA} flowing from contact 1 to 3. 
    (d) Image of the magnetic susceptibility. Bright regions are strongly diamagnetic and therefore superconducting, whereas dark regions are non-superconducting.  Images were acquired simultaneously at \SI{225}{\milli\kelvin}. All scale bars are \SI{25}{\micro\metre}. Part of this data was published in Ref. \cite{Bachmann2019Spatial5}.
    }
    \label{fig:CeIrIn5}
\end{figure}

Fig.~\ref{fig:CeIrIn5} shows an example of measurements on a CeIrIn$_5$ microstructure studied in Ref.\cite{Bachmann2019Spatial5} taken with the scanning SQUID microscope described here. We used a SQUID with a \(\sim\)\SI{1.5}{\micro\metre} sensitive area (pickup loop) and a \(\sim\)\SI{6}{\micro\metre} on-chip field coil which enables local magnetic susceptibility measurements. Applying a current to the field coil applies a small magnetic field to the sample. The SQUID has a gradiometric design, such that the current in the field coil only couples a minimal amount of flux directly into the SQUID \cite{HuberRSI2008}. This amount can be calibrated with the SQUID retracted from the sample. A finite magnetic response by the sample to the field applied with the field coil modifies the flux in the SQUID by an amount that is proportional to magnetic susceptibility of the sample. The device was fabricated from a bulk single crystal of CeIrIn$_5$ using a focused ion beam (FIB). A lamella is cut out from the crystal (outline of the lamella is shown in Fig.~\ref{fig:CeIrIn5}(a) as white dashed line), placed on a substrate, and contacted with a gold layer. The gold is removed in the active area of the device, and four electrical contacts are separated by cutting trenches (black in the image) using a FIB again (see Ref.\cite{Bachmann2019Spatial5} for details). The images were taken at \SI{225}{\milli\kelvin} at which all parts of the CeIrIn$_5$ structure were superconducting. The images shown in Figs.~\ref{fig:CeIrIn5} (b-d) showcase different imaging modes: imaging static stray magnetic fields (Fig.~\ref{fig:CeIrIn5} (b)), imaging the magnetic field produced by an AC current applied to the device (Fig.~\ref{fig:CeIrIn5} (c)), and imaging the local magnetic susceptibility (Fig.~\ref{fig:CeIrIn5} (c)). The static magnetic image ((Fig.~\ref{fig:CeIrIn5} (b)) shows a combination of Meissner screening of a small background field and vortices in the structure. The current was applied from contact 1 to contact 3. The local magnetic susceptibility shows a strong diamagnetic response above the superconducting structure. All images shown in Figs.~\ref{fig:CeIrIn5} (b-d) were acquired simultaneously.

\section{Characterization of relative motion between SQUID and sample}

The pulse tube causes vibrations in the cryostat. These only affect the imaging if they cause a relative motion between the SQUID and sample. We characterize this relative motion by following a method reported in Ref. \cite{SchiesslAPL2016} based on analyzing the excess flux noise caused by vibrations in areas with large magnetic field gradients. 
\begin{figure}[btp]
    \centering
    \includegraphics{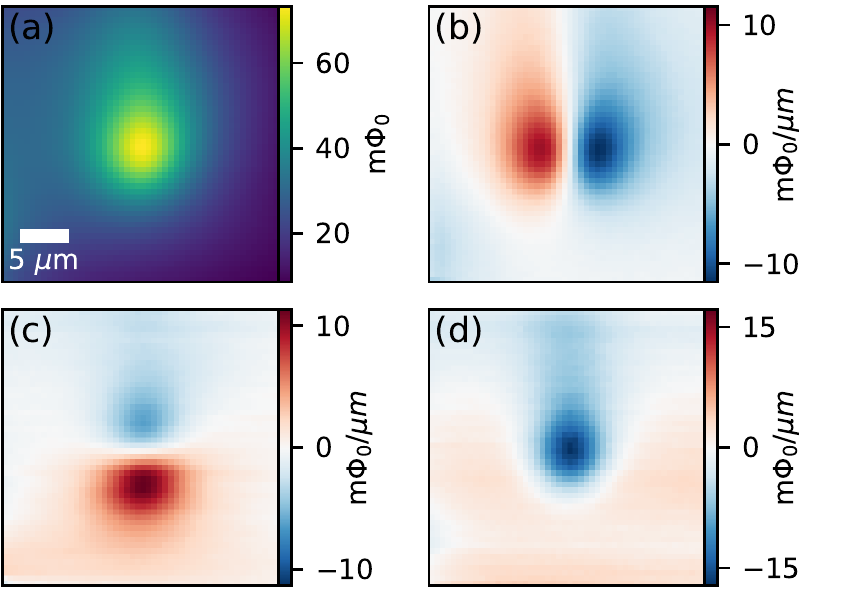}
    \caption{(a) Image of a superconducting vortex in a FIB defined microstructure. ((b-d) Flux gradients in the (b) $x$, (c) $y$, and (d) $z$ directions. The $x$ and $y$ gradients are obtained from numerically differentiating (a), whereas the $z$ gradient is obtained from measurements taken at two heights separated by \(\Delta z = \SI{0.6}{\micro\metre}\).
    }
    \label{fig:deriv}
\end{figure}

\begin{figure}[btp]
    \centering
    \includegraphics{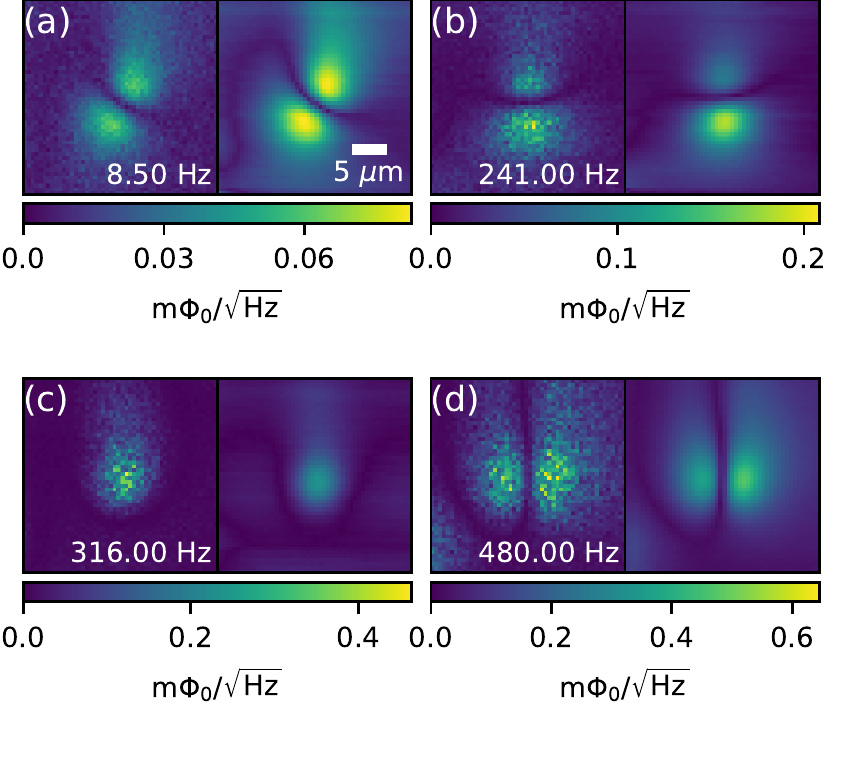}
    \caption{ Spatial maps of the flux noise spectral density (left panel) at the indicated frequencies and corresponding fits to the model in Eq. \ref{eq:model} (right panel). 
    }
    \label{fig:freqimages}
\end{figure}

Fig.~\ref{fig:deriv}(a) shows a vortex in a microstructure fabricated from Sr$_2$RuO$_4$. The corresponding magnetic flux gradients in the \((x,y)\) direction (Figs.~\ref{fig:deriv}(b,c)) are obtained by numerically differentiating Fig.~\ref{fig:deriv}(a). To obtain the flux gradient in the $z$ direction (Fig.~\ref{fig:deriv}(d)) we took a second measurement at $\Delta z = \SI{0.6}{\micro\metre}$ higher than Fig.~\ref{fig:deriv}(a). Relative motion between the SQUID and the sample causes noise in the measured flux signal that depends on the strength and direction of the gradient and the magnitude and direction of the motion. The intrinsic flux noise of the SQUID has a white noise floor on the order of  \SI[per-mode=symbol]{1}{\micro\FluxQuantum\per\sqrthz}  and a $1/f$  tail below ~\SI{100}{Hz}. Given the magnitude of the flux gradients of approximately \SI[per-mode=symbol]{10}{\micro\FluxQuantum\per\micro\metre}, flux noise of \SI[per-mode=symbol]{1}{\micro\FluxQuantum\per\sqrthz} corresponds to an approximate sensitivity of \SI[per-mode=symbol]{0.1}{\nano\metre\per\sqrthz} for detecting relative motion between the SQUID and the sample. Each pixel in Fig.~\ref{fig:deriv}(a) is the average of a four-second time trace. The Fourier transforms of each time trace provide a position-dependent flux noise spectral density. Examples of spatial maps of the noise spectral density at a few frequencies are shown in the left panels of Fig.~\ref{fig:freqimages}.

To find the power spectral density of the vibrations, we model the flux power spectral density at each pixel \((i,j)\) and frequency \(f\) as a sum of vibration-induced noise and intrinsic SQUID and electrical noise \cite{BishopVanRSI2019}:
\begin{widetext}
\begin{align}
    \left[ \hat{\Phi}(f)\right]_{ij}^2 &{}={} 
    \left(
    \left[ \frac{\partial \Phi }{\partial x} \right]_{ij}\mkern-5mu
    \hat{X}(f) + 
    \left[ \frac{\partial \Phi }{\partial y} \right]_{ij}\!\!
    \hat{Y}(f) + 
    \left[ \frac{\partial \Phi }{\partial z} \right]_{ij}\!\!
    \hat{Z}(f)\right)^2 
    + \left(\hat{N}(f)\right)^2. \label{eq:model}
\end{align}
\end{widetext}
Here, \(\hat{\Phi}(f)\) denotes the flux noise amplitude spectral density, \(\partial \Phi/\partial(x,y,z)\) are the flux gradients shown in Fig.~\ref{fig:deriv}, \(\hat{X}(f),\hat{Y}(f),\hat{Z}(f)\) are the vibrations along the (\(x,y,z\)) direction, and \(\hat{N}(f)\) models intrinsic SQUID noise and is strictly positive. We fit the spatial map of the noise to Eqn.~\eqref{eq:model} to obtain the spatial vibrations \(\hat{X},\hat{Y},\hat{Z}\) and electrical noise \(\hat{N}\) at each frequency. We use statistical bootstrapping with 200 trials to avoid local optima and determine the 95\% confidence intervals of the fits \cite{EfronSS1986}. 

In the right panels of Fig.~\ref{fig:freqimages}(a-d) we show examples of fits to the maps of flux noise spectral density. Figs.~\ref{fig:freqimages}(b,c,d) closely resemble the gradients in the (\(y,z,x\)) directions respectively as shown in Fig.~\ref{fig:deriv}. This suggests that the vibrations at these frequencies are mostly in a single Cartesian direction. Fig.~\ref{fig:freqimages}(a) shows an example of motion along both the \(x\) and \(y\) direction.  

\begin{figure*}[btp]
    \centering
    \includegraphics{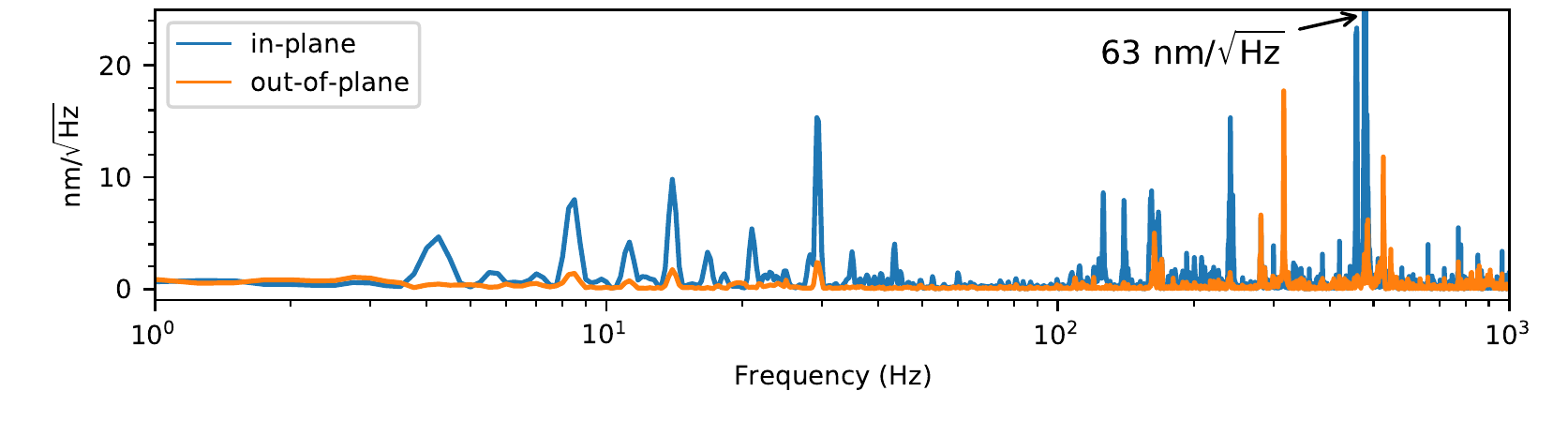}
    \caption{ Vibration spectral densities for in-plane (blue) and out-of-plane (orange) relative motion of the SQUID and sample. Measurements were taken with the air spring pressurized and the mixing chamber temperature at \SI{80}{\milli\kelvin} with the \He{3}/\He{4} mixture circulating. The vertical axis ranges to \SI[per-mode=symbol]{25}{\nmnoise} cutting off the highest peak that reaches \SI[per-mode=symbol]{63}{\nmnoise} at \(\sim\) \SI{480}{Hz} in the in-plane vibration spectrum.
    }
    \label{fig:spectrum}
\end{figure*}

 Fig.~\ref{fig:spectrum} shows the vibration spectral densities for in-plane and out-of-plane directions obtained by fitting at each frequency. Error bars estimated from the 95\% confidence intervals of the fits are less than \SI[per-mode=symbol]{1}{\nmnoise} at all frequencies and omitted to improve readability. Systematic uncertainty comes from uncertainty in the conversion of voltage applied to the piezoelectric scanners to the induced displacement in $\mu$m. This conversion is only used in the final step of the analysis, since all measurements, fits and computations are carried out in terms of voltages applied to the piezoelectric scanners. We determine the conversion factors for motion in $x$ and $y$ as 175$\pm$10 nm/V from imaging lithographically-defined samples with known dimensions and features. To obtain the conversion factor for vertical motion we analyze a series of images of a vortex taken at different heights. Using the known height dependence of the magnetic field profile of the vortex, we estimate the conversion to be $149\pm 6$ nm/V.

Pressurizing the room-temperature air springs that float the top plate of the cryostat has the most notable impact on the sample-to-SQUID vibrations. The air springs significantly suppress vibrations below \SI{200}{Hz}. However, we find a surprising amount of motion in a frequency band between \SI{450}{Hz} and \SI{490}{Hz}. The amplitude of the vibrations in this range vary significantly in amplitude over long time scales. However, we could not correlate their behavior with any changes in the cryostat or the lab environment. From the structure of the noise spectral density maps, we can determine that the most pronounced peak  at \(\sim\) \SI{480}{Hz} corresponds to motion along the $x$-direction (see for example Fig.~\ref{fig:freqimages}(d)). Fig.~\ref{fig:spectrum} is based on a dataset in which these vibrations are particularly pronounced. The integrated in-plane (out-of-plane) vibrations are \SI{31}{nm} (\SI{27}{nm}) from \SI{0.25}{Hz} to \SI{450}{Hz} compared to \SI{106}{nm} (\SI{39}{nm}) from \SI{0.25}{Hz} to \SI{1000}{Hz}. With the air springs turned off, we observed integrated in-plane (out-of-plane) vibrations of \SI{90}{nm} (\SI{32}{nm}) from \SI{0.25}{Hz} to \SI{450}{Hz}.

We did not observe a noticeable difference between the vibrations with and without the circulation of the mixture running. We have recorded motion of the mixing chamber plate along the vertical direction using a geophone, since we cannot measure the sample-to-SQUID vibrations with the pulse tube turned off. With the vacuum cans closed and the fridge at room temperature, no noticeable motion is detected with the pulse tube turned off, however with the pulse tube turned on, some motion is clearly present at higher frequencies. However the amplitudes and exact positions of peaks are not strongly correlated with the peaks visible in Fig.~\ref{fig:spectrum}. We speculate that the motion we observe above \SI{200}{Hz} has contributions from the pulse tube and that resonances in the microscope shape the spectrum we observe, but that an additional source of vibrations is intermittently present in the lab.

\section{Conclusions}
We demonstrate the operation of a scanning SQUID microscope with several imaging modes in a cryogen-free DR. We avoid the use of a spring stage and designed a rigid microscope housing to optimize for thermal anchoring of the sample. Our microscope allows for a large number of wires and coaxial connections and fits in the bore of a superconducting magnet. The temperature of the sample stage reaches at least \SI{30}{\milli\kelvin} and does not increase significantly during scanning. In Ref. \cite{BishopVanRSI2019} the authors report the relative sensor-to-sample vibrations in a cryogen-free Bluefors cryostat with a \SI{2.8}{\kelvin} base temperature with and without using a spring stage. Our performance falls in between these two benchmarks. We believe that this is achieved due to a combination of a more rigid construction of the microscope and lower vibrations in the cryostat itself. In the future, we hope to further reduce the vibrations through a combination of optimizing the coarse and fine positioning assembly for stiffness, the thermal braids that connect the pulse tube and cold plates, and the connection between the motor valve and the cold head. We plan to use this microscope to study novel superconductors, topological phases of matter, and frustrated magnetic systems down to millikelvin temperatures. 

\section{Acknowledgements}
We thank Eric Smith, Jihoon Kim and Kevin Nangoi for help with the construction of the microscope, Bluefors for technical support and discussions, and Eric D. Bauer, Filip Ronning, Naoki Kikugawa, Andrew P. Mackenzie for growing and providing the crystals for the microstructures. This work was supported by the U.S. Department of Energy, Office of Basic Energy Sciences, Division of Materials Sciences and Engineering, under award DE-SC0015947 (scanning SQUID imaging, implementation of millikelvin microscope) and the Cornell Center of Materials Research with funding from the NSF MRSEC program under award DMR-1719875 (SQUID and microscope design). Fabrication of the microstructures was supported by the Max Planck Society and by the Deutsche Forschungsgemeinschaft (DFG, German Research Foundation) – MO 3077/1-1 and the European Research Council (ERC) under the European Union’s Horizon 2020 research and innovation program (GA 715730).

\section{Data Availability}
The data that support the findings of this study are available from the corresponding author upon reasonable request.  The drawings of the microscope are available at \cite{CADDrawing}.

\bibliography{References_CryogenFreeDR,References_from_Mendeley}
\end{document}